\documentclass[fleqn,usenatbib]{mnras}

\usepackage{mathptmx}
\usepackage{txfonts}
\usepackage{textcomp}
\usepackage{ulem}

\usepackage[T1]{fontenc}
\usepackage{ae,aecompl}

\usepackage{graphicx}	
\usepackage{amssymb}	
\usepackage{natbib}
\usepackage{graphicx} 
\usepackage{hyperref}
\usepackage{txfonts}
\usepackage{umoline}

\title[Testing the third body hypothesis in the four Cataclysmic Variables]{Testing the third body hypothesis in the Cataclysmic Variables {\sl LU Camelopardalis}, {\sl QZ Serpentis}, {\sl V1007 Herculis} and {\sl BK Lyncis.}}

\author[C. Chavez et al.]{
Carlos E. Chavez,$^{1}$\thanks{E-mail: Carlos.ChavezPCH@uanl.edu.mx}
Nikolaos Georgakarakos,$^{2,3}$
Andres Aviles,$^{4}$
Hector Aceves,$^{5}$
\and Gagik Tovmassian,$^{5}$
Sergey Zharikov,$^{5}$
J. E. Perez--Leon$^{4}$
and Francisco Tamayo$^{4}$
\\
$^{1}$Universidad Auton\'oma de Nuevo Le\'on, Facultad de Ingenier\'{i}a Mec\'anica y El\'ectrica, San Nicol\'as de los Garza 66451, NL, M\'exico\\
$^{2}$New York University Abu Dhabi, P.O. Box 129188, Saadiyat Island, Abu Dhabi, UAE\\
$^{3}$Center for Astro, Particle and Planetary Physics (CAP3), New York University Abu Dhabi, P.O. Box 129188,
Saadiyat Island,\\ Abu Dhabi, UAE\\
$^{4}$Universidad Auton\'oma de Nuevo Le\'on, Facultad de Ciencias F\'{i}sico--Matem\'aticas, San Nicol\'as de los Garza 66451, NL, M\'exico\\
$^{5}$Universidad Nacional Aut\'onoma de M\'exico, Instituto de Astronom\'\i{}a, Ensenada 22860, BC, M\'exico\\
}

\date{Accepted XXX. Received YYY; in original form ZZZ}

\pubyear{2022}

\begin{document}
\label{firstpage}
\pagerange{\pageref{firstpage}--\pageref{lastpage}}
\maketitle

\begin{abstract}
Some Cataclysmic Variables (CVs) exhibits a very long photometric period (VLPP). We calculate the properties of a hypothetical third body, initially assumed on circular--planar orbit, by matching the modelled VLPP to the observed one of four CVs studied here: {\sl LU Camelopardalis} ({\sl LU Cam}), {\sl QZ Serpentis} ({\sl QZ Ser}), {\sl V1007 Herculis} ({\sl V1007 Her}) and {\sl BK Lyncis} ({\sl BK Lyn}). 
 The eccentric and low inclination orbits for a third body are considered using analytical results.
 The chosen parameters of the binary components are based on the orbital period of each CV. The smallest corresponding semi-major axis permitted before the third body\textquotesingle s  orbit becomes unstable is also calculated. 
A first-order analytical post-Newtonian correction is applied, and the rate of precession of the pericentre is found, but it can not explain any of the observed VLPP. 
For the first time, we also estimate the effect of secular perturbations by this hypothetical third body on the mass transfer rate of such CVs. We made sure that the observed and calculated amplitude of variability was comparable too. 
The mass of the third body satisfying all constrains range from 0.63 to 97 Jupiter masses. 

Our results show further evidence supporting the hypothesis of a third body in three of these CVs, but only marginally in {\sl V1007 Her}.

\end{abstract}




\section{Introduction}

A Cataclysmic Variable (CV hereinafter) is a semidetached binary star system that is particularly stable (Frank et al. 2002). A CV consists of a white dwarf (WD) primary star and a lower mass main-sequence secondary star, mainly a M star, although the spectral class can range from a K to L type star. The condition that defines the distance between the two components is that the main sequence star fills its corresponding Roche lobe and loses matter through the $L_1$ Lagrangian point. The matter accretes onto the WD via an accretion disc, unless the WD has a strong enough magnetic field to prevent the formation of the disc. If the system is disturbed for any reason, it tends to restore to equilibrium. 

Two mechanisms maintain the balance of a CV: the evolutionary expansion of the secondary, and the decrease of the semi-major axis of the binary due to the loss of angular momentum. The decrease in angular momentum has two possible sources, depending on the orbital period of the binary system. The first one is magnetic braking, and it is the dominant effect for systems that have orbital periods above three hours (Whyte and Eggleton 1985,Livio and Pringle 1994). The second one is the emission of gravitational waves, for systems that have orbital periods below two hours (Faulkner 1976; Chau and Lauterborn 1977). In the 2-3~h period range, neither mechanism is efficient for the angular momentum removal. Hence, the number of known systems in that range, called the period gap, is significantly smaller.

The material that the secondary loses through the $L_1$ point can not fall straight to the primary; instead, it forms an accretion disc (Frank et al. 2002; Ritter 2008). This accretion disc is so luminous that it outshines both stars. The disc brightness is proportional to the mass transfer rate. Therefore, if the mass transfer rate changes, the luminosity of the system changes.  In particular,  a change in the location of the $L_1$ point will change, the mass transfer rate, and as a consequence, the luminosity of the whole system will change.  

CVs are notorious for variability on different time and magnitude scales.  In this paper, we are going to consider a specific one: a relatively low amplitude (0.07--0.97~mag) variability with periods exceeding the orbital ones hundreds to thousands of times. The very long photometric period (VLPP, hereinafter) was first singled out in {\sl FS Aur} (Chavez et al. 2012, 2020), although the object shows many other variabilities. More VLPPs have been identified in other CVs  (e.g. Thomas et al. 2010; Kalomeni 2012; Chavez et al. 2012; Yang et al. 2017; Chavez et al. 2020).

Different mechanisms have been proposed to explain VLPPs. Thomas et al. (2010) found a long-term modulation with a period of 4.43 days in the CV {\sl PX And}, using eclipse analysis, and proposed the disc precession period as the origin of the VLPP. Another example is the {\sl DP Leo} system, where Beuermann et al. (2011) found a period of 2.8 yr, using eclipse time variations, and concluding that a third body was the best explanation for the VLPP. Honeycutt, Kafka and Robertson (2014) found a 25--d periodicity in {\sl V794}. Kalomeni~(2012) discovered several magnetic CVs that have long--term variability, with a time scale of hundreds of days and concluded that those VLPPs are likely to originate from the modulation of mass--transfer due to the magnetic cycles in the companion star.

More recently, Chavez~et~al. (2012, 2020), using dynamical analysis, proposed that a third body can induce a VLPP by secular perturbations on the inner binary. The third body can introduce oscillations of the $L_1$ point of the close binary and, therefore, the mass transfer rate changes. This mechanism induces periods as long as the VLPPs in the inner binary by means of secular perturbations observed in the above mentioned  CVs. 
The VLPP  was observed in the long--term light curves of ten CVs by  Yang~et~al.~(2017). As a possible mechanism of the VLPP for five out of ten systems, these authors proposed a third body orbiting the close binary, with the system being in Kozai--Lidov resonance (Kozai 1962, Lidov 1962) which requires an orbital inclination between the plane of the binary and the orbit of the third object larger than 39.2$^{\circ}$. By that, they were able to estimate the possible orbital period of the third body.

Our main goal in this research is to investigate whether a third body can explain the observed VLPP of the four CVs, rather than obtaining a precise value on the mass of the third body.

This paper is organized as follows. Section~2 provides information about the CVs considered in this work, and their initial parameters. Section~3 gives the properties of the third body that result from our analysis in order to explain the observed VLPPs. Section~4 briefly addresses the potential role of a post-Newtonian correction on the VLPPs. In Section~5, we address the effects of a probable third body on the mass transfer rate  and brightness of the four CVs. In section~6, we presents our results and discussion, and we provide final comments on this work in Section~7.


\section{THE CATACLYSMIC VARIABLES STUDIED AND THEIR INITIAL PARAMETERS}

\label{Sec:CVs}

Yang~et~al.~(2017) matched 344 out of 1580 known CVs, and extracted their data from the Palomar Transient Factory (PTF) data repository. These images were combined with the Catalina Real-Time Transit Survey (CRTS) light curves. They found ten systems with unknown VLPPs; {\sl BK Lyncis} (2MASS J09201119+3356423), {\sl CT Bootis}, {\sl LU Camelopardalis} (2MASS J05581789+6753459), {\sl QZ Serpentis} (SDSS J155654,47+210719.0), {\sl V825 Herculis} (2MASS J17183699+4115511), {\sl V1007 Herculis} (1RXS J172405.7+411402), {\sl Ursa Majoris 01} (2MASS J09193569+5028261), {\sl Coronae Borealis 06} (2MASS J15321369+3701046), {\sl Herculis 12} (SDSS J155037.27+405440.0) and {\sl VW Coronae Borealis} (USNO-B1.0 1231-00276740).
They analyse each system and depending on the value of its VLPP propose a most likely origin, such as the precession of the accretion disc, hierarchical three--body systems and magnetic field change of the companion star. They argue that if the long--term period is less than several tens of days, the disc precession explanation is preferred. However, the hierarchical three body system or the variations in the magnetic field are favoured for longer periods.
Six out of those ten systems they propose to be a hierarchical triple: {\sl BK Lyn}, {\sl LU Cam}, {\sl QZ Ser}, {\sl V1007 Her}, {\sl Her 12} and {\sl UMa 01}.

{\sl UMa 01}, has a long orbital period of $P_{1}=404.10 \pm 0.30$~min (6.735~h); long compared with other systems in the sample. According to the orbital period distribution for CVs, the number of systems with such period or larger is very small and therefore most of the statistical results cannot be applied to them (Knigge 2006; Knigge, Baraffe and Patterson 2011). {\sl UMa 01} has been presumably formed recently and there are not good enough estimates of its parameters (e.g. mass, radius, temperature) of either component.
Additionally, {\sl Her 12} was identified as a CV by Adelman-McCarthy~et~al.~(2006), but we do not model it since its period is not well constrained, possibly being in the range between $P_{1}=76 $--$174$~min (Yang et al. 2017).
We study each of the four systems remaining (i.e.{\sl LU Cam}, {\sl QZ Ser}, {\sl V1007 Her} and {\sl BK Lyn}) to learn more about their dynamical attributes.

The values reported by Knigge~et~al.~(2011) are used for calculating the mass of each member of the CV. We did so since in their article they give all the parameters that we later use in this research such as mass, radius, semi-major axis for each component of the binary. In that study they used eclipsing CVs and theoretical restrictions to obtain semi--empirical donor sequence for CVs with orbital periods $P_{1}<6$~h. They estimate all key physical, photometric and spectral--type parameters of the secondary and primary as a function of the orbital period. 
We use the data from their Table~6 and 8  (Knigge~et~al.~2011) to obtain the parameters of the CV\textquotesingle s in our selection. In practice, we use the online version of those tables (that are far more complete) to obtain the adequate values for the CVs studied here. If the systems have any peculiar features we will point it out in the text and we will state the reference used for such a value.

\subsubsection{White Dwarf mass}

First, we briefly describe the mass value used by Knigge~et~al.~(2011). In their research they explain that they used the mean value of $\langle M_{1}\rangle= 0.75 \pm 0.05 \, {\rm M_{\odot}}$. In 2011 the new data pointed to a mean value for the WD in CV\textquotesingle s of $\langle M_{1} \rangle= 0.79 \pm 0.05 \, {\rm M_{\odot}}$. They stated that since they had already begun to assemble the grid of donors sequence and evolution tracks ``we chose to retain  $\langle M_{1}\rangle= 0.75 \pm 0.05 \, {\rm M_{\odot}}$ as a representative of WD mass''. More recently, in a review by Zorotovic and Schreiber~(2020), it was reported that the mean WD value could be even higher, between $\langle M_{1}\rangle= 0.82-0.83 \pm 0.05 \, {\rm M_{\odot}}$.
   We decided to use the  Knigge~et~al.~(2011) values for all parameters of the WD to be self--consistent throughout this article and also because they provide estimates for $M_{1}$ and  $R_{1}$ (WD\textquotesingle s mass and radius) corresponding to the orbital period of each CV (both values necessary for the calculations of the next sections).

 To understand how this affects the calculations we would like to point out that back in Chavez~et~al.~(2012), the calculations were done with $M_{1}=0.7 \, {\rm M_{\odot}}$ and then we updated it to $M_{1}=0.75 \,{\rm M_{\odot}}$ (a change of 7\%) in  Chavez~et~al.~(2020). 
   The minimum in the plot of  Fig.~8 (2012 article) middle panel (semi--major axis vs mass of the third body) has a value of $M_{3}=50\, {\rm M_{J}}$, while when $M_{1}=0.75\,  {\rm M_{\odot}}$ is used (Fig.~3, 2020 article) the minimum corresponds to $M_{3}=30\,  {\rm M_{J}}$. 
  That is a 40\% decrease of the mass of the third body at the minimum.

\subsubsection{LU Camelopardalis}

{\sl LU Cam} is a dwarf nova CV and the first spectrum of this system was obtained by Jiang~et~al.~(2000). 
Its orbital period was first reported by Sheets~et~al.~(2007) to be $P_{1}=0.1499686(7)$~days$=3.599246$~hr. There, they point out that the averaged spectrum shows a strong blue continuum. Yang~et~al.~(2017) report a VLPP of 265.76 days and point out the hierarchical triple system explanation as their best candidate to explain it.

Using data from Knigge~et~al.~(2011) we obtain $M_{1}=0.75 \, {\rm M_{\odot}}$, $M_{2}=0.26 \,  {\rm M_{\odot}}$.  We show all the parameters of the system in Table \ref{tab:initial}.

\subsubsection{QZ Serpentis}

{\sl QZ Ser} is a system that has been classified as a dwarf nova. The system has an orbital period of  $P_{1}=119.752(2)$~min $=1.99584$~h according to Thorstensen~et~al.~(2002a). 
 These authors found that the system is not a usual CV,  as it is one of a few objects known with a short orbital period and a secondary non--standard K-type star. This K-type secondary has a much smaller mass than a usual K star because of unstable thermal scale mass transfer evolution. There are other examples of this type of CVs. For instance, Thorstensen~et~al.~(2002b) found a K4 in the dwarf nova 1RXS J232953.9+062814, while Ashley~et~al.~(2020) found a K5 around a CV with a period of 4.99~h.

Thorstensen~et~al.~(2002a) used evolutionary models to estimate {\sl QZ Serpentis} parameters such as $M_{2}=0.125 \pm 0.025 \, {\rm M_{\odot}}$ which yielded $R_{2}=0.185 \pm 0.013 \, {\rm R_{\odot}}$, where $R_{2}$ is the secondary\textquotesingle s radius. They also used a typical white dwarf mass value of $M_{1}=0.7 \, {\rm M_{\odot}}$, widely used in 2002 (Jiang et al. 2000; Thorstensen at al. 2002a).

Thorstensen~et~al.~(2002a) estimated from observations of the ellipsoidal variations that the inclination (with respect to sky\textquotesingle s plane) of the system must be $i=33.7^{\circ} \pm 4^{\circ}$.
Then decided to use this estimate to constrain the secondary's mass. They proceeded to check mass ratios between the primary and the secondary between 0.1 to 0.4 for this system.  Using the secondary's velocity amplitude they give a mass function of $f=0.075 (5) \, {\rm M_{\odot}}$. The inclination can be calculated from the masses and the mass function using the following equation:
\begin{equation}
\label{incl}
 i =  \arcsin \Bigg[ \bigg(  {(M_{1}+M_{2})^2 f  \over M_{1}^3}  \bigg) ^ {1 \over 3} \Bigg].
\end{equation}
Taking $M_{2}=0.125 \, {\rm M_{\odot}}$ and $M_{1}=0.7 \,  {\rm M_{\odot}}$,   Thorstensen et~al.~(2002a) obtained a value of $i=32^{\circ}$.

If we calculate the statistical values obtained by Knigge~et~al.~(2011), for the parameters of this CV (using the orbital period to do so) we find that $M_{1}=0.75 \,  {\rm M_{\odot}}$, $R_{1}=0.0107 \, {\rm R_{\odot}}$, $M_{2}=0.15 \, {\rm M_{\odot}}$ and $R_{2}=0.1923 \, {\rm R_{\odot}}$. Therefore, these $M_{2}$ and $R_{2}$ estimates are both well within the uncertainties of the estimates of 
Thorstensen~et~al.~(2002a). As we pointed out earlier, we decided to use the Knigge~et~al.~(2011) values to be self--consistent throughout this article since we need estimates for $M_{1}$ and $R_{1}$; both values will be used in the following sections.

Additionally, using these values in Eq.~\ref{incl} we obtain a value for the inclination $i=31.7^{\circ}$, which is well within the observational inclination uncertainty estimated by Thorstensen~et~al.~(2002a) and very close to the value they provide.

The VLPP found by Yang~et~al.~(2017) is 277.72, which is the longest among the four systems studied, and conclude that a hierarchical triple system is the best scenario that can explain this period. Table \ref{tab:initial} shows the parameters used for this system in this work.

\subsubsection{V1007 Herculis}

This CV was discovered by Greiner~et~al.~(1998). They found that it is a polar system with an orbital period of $P_{1}=404.10 \pm 0.30$~min $=1.9988$~hr. Since it is a polar system there is no disc around it, and there are no periods associated with the disc. Greiner~et~al.~(1998) estimated the mass of the secondary using the orbital period and found it to be $M_{2}=0.16 \, {\rm M_{\odot}}$; to do so they assumed a mass--radius relationship for main sequence stars using Patterson~(1984).

Using the parameters of Knigge~et~al.~(2011) for this CV we obtain that $M_{1}=0.75 \, M_{\odot}$ and $M_{2}=0.15\,  M_{\odot}$, also shown in Table \ref{tab:initial} along with the rest of the parameters. The observed VLPP  by Yang~et~al.~(2017) is 170.59 days.

\subsubsection{BK Lyncis}

{\sl BK Lyn} is a nova--like CV which was discovered by Green~et~al.~(1998). The calculated orbital period is $P_{1}=107.97 \pm 0.07$ min$=1.7995$ h, found by Ringwald et~al.~(1996). In addition, the secondary was found to be a M5V star by using infrared spectroscopy by Dhillon et~al.~(2000). The accretion rate was found to be between $\dot{M}_{WD} \approx 10^{-8}$--$10^{-9} \,\textrm{M}_{\odot}/\textrm{yr}$, constraining the mass of the WD in a wide range of values between 0.4$\,{\rm M}_\odot$ and 1.2$\,{\rm M}_\odot$. Yang~et~al.~(2017) found that the VLPP for this system is 42.05 days (the lowest among all CVs studied here) and ruled out other possible explanations except for a hierarchical triple system one.

Using Knigge~et~al.~(2011), as pointed out in the previous subsection, we obtain $M_{1}=0.75\, {\rm M}_{\odot}$ and $M_{2}=0.13 \, {\rm M}_{\odot}$, with all the parameters of the system shown in Table~\ref{tab:initial}

\begin{table*}
\caption{\label{tab:initial} Initial parameters and magnitudes for all systems are calculated using Knigge~et~al.~(2011). The observed minimum magnitude ($M_{Bmin}$), maximum ($M_{Bmax}$) and overall change ($\Delta M_{B}$) due to VLPP (Yang~et~al.~2017) are shown.}
\begin{tabular}{@{}lcclrrlllllr}
\hline
 Name of the CV& Binary Period  & $M_{1}$ & $M_{2}$ &  $R_1$  & VLPP &  $M_{2} / M_{1}$ &   $a$ & $M_{Bmax}$ & $M_{Bmin}$ & $\Delta M_{B}$ & $\log (\dot{M}_{2})$  \\
  & (hours) & (${{\rm M}_{\odot}}$)  & (${{\rm M}_{\odot}}$) &  (${\rm R}_{\odot}$)  & (days) & & (AU) & & & &    $\   \ ({{\rm M}_{\odot}/{\rm yr} })$\\
 \hline

 \hline
{\sl LU Camelopardalis} & 3.5992  & 0.75 & 0.26 & 0.011 & 265.76 & 0.34 & 0.0055 & 15.55 & 16.10 & 0.55 &-9.02 \\
{\sl QZ Serpentis} & 1.99584  & 0.75 & 0.15 & 0.011 & 277.72 & 0.20 & 0.0036 & 17.43 & 17.50 & 0.07 & -10.09\\
{\sl V1007 Her} & 1.99883  & 0.75 & 0.15 & 0.011 & 170.59 & 0.20 & 0.0036 & 17.83 & 18.80 & 0.97 & -10.09\\
{\sl BK Lyncis} & 1.7995  & 0.75 & 0.13 & 0.011& 42.05 & 0.17 & 0.0033 & 14.40 & 15.08 & 0.68 & -10.14 \\
\hline
\end{tabular} \\
\end{table*}

\section{THREE--BODY CATACLYSMIC VARIABLE}

As pointed out earlier, Yang~et~al.~(2017) proposed the hierarchical triple system hypothesis for the four systems studied here after ruling out other explanations. There, they explored the Lidov--Kozai resonances as a possible explanation for the VLPP observed, and found the possible semi--major axis of the third body. The mutual inclination between the inner binary orbital plane and the third--body orbital plane should be greater than $39.2^{\circ}$ for this mechanism to be effective in disturbing the inner binary effectively.  

Here we explore a new possibility, namely that the secular perturbation by a low eccentricity and low inclination third object explains the VLPP and also the change of magnitude observed in these four CVs.

\subsection{Third body on a close near--circular planar orbit} 

Chavez~et~al.~(2012), while investigating the system {\sl FS Aurigae}, ruled out that the VLPP could correspond directly to the period of a third body, since the object would be too distant to have an important effect on the inner binary. A series of numerical integrations were performed and showed that indeed the effect is minimal  and could not explain the VLPP of the CV {\sl FS Aurigae}.

It was concluded that a third body on a close near--circular planar orbit could produce perturbations on the central binary eccentricity, and they are modulated at three different scales, the period of the binary $P_1$, the period of the perturber $P_2$ and the much longer secular period--VLPP. Secular perturbations have been studied both analytically and numerically by Georgakarakos~(2002, 2003, 2004, 2006, 2009). A third body prevents the complete circularization of the orbit due to tides by producing a long--term eccentricity modulation (e.g. Mazeh~\&~Shaham~1979; Soderhjelm~1982, Soderhjelm~1984, Chavez~et~al~2012, 2020). From Georgakarakos~(2003) it is possible to estimate the amplitude of such eccentricity by using the following equation: 
 \begin{equation}
\label{deltae}
\Delta e_{1} \propto q_{3} \Big(  {P_{1} \over P_{2}} \Big)^{8 / 3}  e_{2}  \big( 1- e_{2}^2 \big)^{-5 / 2},
\end{equation}
where $P_{2}$ is the period of the third body around the inner binary, $e_{2}$ is the eccentricity of the orbit and $q_{3}=M_{3}/(M_{1}+M_{2}+M_{3})$. Therefore any changes over time on the eccentricity $e_{2}$, such as the modulations studied in Chavez~et~al.~(2012), will have an effect on the eccentricity $e_{1}$ of the CV, modulating and changing the position of the $L_1$ point and hence changing the brightness of the system. The details of the numerical modelling will be given in the next section.

\subsection{Numerical modelling for the circular case} 

We performed dynamical simulations of the CVs with a hypothetical third body. The high--order Runge--Kutta--Nystrom RKN 12(10)~17M integrator of Brankin~et~al.~(1989)  was used for the equations of motion of the complete three body problem in the barycentre inertial reference frame. The total energy was conserved to $10^{-5}$ or better for all numerical experiments.

As in Chavez~et~al.~(2012), tidal deformation of the stars in the close binary is not important for CVs in general and the two objects can be considered point masses. Hence, all three bodies are considered point masses in our integrations. The binary is initially on a circular orbit, and the third mass moves initially on its own circular orbit around the inner binary in the same plane. The mass $M_3$ and its orbital period $P_2$ are chosen across an ensemble of numerical experiments.

We proceed as follows. We fix the value of the period of the third body $P_2$, we change its mass $M_3$, we perform the numerical integrations, and then the eccentricity $e_1$ is calculated as a function of time. We obtain the secular period on each integration from $e_1$ using a Lomb--Scargle periodogram (Lomb~1976, Scargle~1982). All this shows the effect that the mass has on the secular period.

In Figs.~\ref{fig:LUCamel}--\ref{fig:BKLyn} we show, as a function of mass ,the VLPPs and semi--major axis obtained from our numerical experiments for each of our CVs studied. Each curve represents a given $P_2$ period that remains constant as we change the mass. We joined the points by using an interpolated curve (spline method) on each case. A black point that appears, for example, in Figure~\ref{fig:LUCamel} middle panel represents a system that can explain the observed VLPP; i.e., any given point represents a combination of semi--major axis and mass that can produce by secular perturbations the observed VLPP.

\subsection{Analytical modelling of the third body on an eccentric and inclined orbit} 
Following Chavez~et~al.~(2020), we also investigate the effect that eccentricity and inclination of the third body may have on the resulting VLPP and the expected parameters of mass and semi-major axis of the third body.

We decided to use previously derived analytical results to see the effect of eccentricity and inclination. The orbital evolution of hierarchical triple systems has been studied in a succession of articles (Georgakarakos~2002, 2003, 2004, 2006, 2009, 2013, and Georgakarakos, Dobbs-Dixon \& Way,~2016).
   Part of these studies was focused on the secular evolution of such systems. These analytical results can give us estimates about the inner binary\textquotesingle s frequency and period of motion. Hence, we can determine which mass values and orbital configurations of a potential third body companion can give rise to the secular periods observed in each CV.

We use the results of Georgakarakos~(2009) for a coplanar perturber on a low eccentricity orbit, and for coplanar systems with eccentric perturbers we make use of Georgakarakos~(2003). Finally, for systems with low eccentricity and low mutual inclinations (with $i_m<$39.23$^{\circ}$) the results of Georgakarakos~(2004) are used.

The analytical expressions for the frequencies and periods can be found in the appendix of this article, while details of derivations can be found in the articles mentioned above.

In Figs.~\ref{fig:LUCamel}--\ref{fig:BKLyn} we show the analytical estimates as curves in different colors depending on the third\textquotesingle s body initial eccentricity or inclination.

\section{Effect of Post--Newtonian correction}

Here, we also consider other dynamical effects that may produce the long term signal we observe in the light curve of the stellar binaries.
We study the effect of a first order post--Newtonian general relativity (GR) correction to the orbit of the stellar binary.\\
For all stellar pairs under investigation, the small semi-major axis of the orbit makes it an interesting case to include a post-Newtonian correction to describe the system\textquotesingle s motion more accurately.  Inclusion of a post-Newtonian correction to our orbit produces an additional precession of the pericentre at the following rate (e.g. Naoz~et~al.~2013, Georgakarakos~\&~Eggl~2015):
\begin{equation}
\label{eqverpi}
\dot{\varpi}=\frac{3 {G}^{\frac{3}{2}}(M_1+M_2)^{\frac{3}{2}}}{c^2a^{\frac{5}{2}}_1(1-e^2_1)},
\end{equation}
where $G$ is the gravitational constant, $c$ is the speed of light in vacuum, $a_1$ is the semi--major axis of the inner binary and $e_1$ the eccentricity of the inner binary.
Based on the precession rate given in the above equation, the post--Newtonian pericentre circulation period for all systems is shown in Table~\ref{tab:GR}.  periods calculated are too long to explain any of the VLPPs.

\section{Effect of the third body on the mass transfer rate and brightness} 

\subsection{Non--Magnetic cases} 

It is possible to estimate how the modulation of the inner binary, due to the secular perturbation of the third body, affects the mass transfer and the brightness of the system. First we focus our attention in the non-magnetic cases, that is {\sl LU Cam}, {\sl QZ Ser} and {\sl BK Lyn}. This subsection follows Chavez~et~al.~(2020), and a brief review is provided here.

To calculate the mass loss of the secondary it is necessary to make use of the definition of $R_{L}(2)$. Calculating directly the volume of the Roche lobe is difficult, so it is better to define an equivalent radius of the Roche lobe    
as the radius, $R_{L}(2)$, of a sphere with the same volume as the Roche lobe. 
Sepinsky~et~al.~(2007) generalized the definition of $R_{L}(2)$ including eccentric binaries, as:
\begin{equation}
\label{RL2}
R_L(2)=r_{12}(t) \; {{0.49 q^{2/3}} \over {0.6 q^{2/3} + \ln{(1+q^{1/3})}}},
\end{equation}
where $r_{12}$ is the distance between the two stars at any given time. We can obtain  $r_{12}$ from our numerical integrations for each system.

Now we want to know the change in magnitude that produces that particular combination of parameters, and then we can compare with the observed magnitude change in the light curve. Therefore, we can find the system in each case that better explains observations according to our calculations.

We proceed as follows to estimate the change in magnitude due to the previous choice of parameters. We can calculate the maximum $R_L(2)_{max}$, shown as a blue horizontal line in Fig.~\ref{fig:LUCamelexplain} and the minimum $R_L(2)_{min}$, shown as a red horizontal line in Fig.~\ref{fig:LUCamelexplain} for each system directly form our numerical results. From here, we can estimate the mass transfer rate $\dot{M} (2)$ and hence the value of the luminosity of each CV.

Assuming that the secondary is a polytrope of index 3/2 and that the density around $L_1$ is decaying exponentially, it is possible to estimate the mass transfer rate using Eq.~2.12 of Warner~(1995):
\begin{equation}
\label{dotM2}
\dot{M}(2)= - C {M(2) \over P_{1}}   \Bigg({\Delta R \over {R(2)} } \Bigg)^{3},
\end{equation}
where $C$ is a dimensionless constant $\approx 10-20$, $R(2)$ is the secondary stellar radius and $\Delta R$ is the amount by which the secondary overfills its Roche Lobe: $\Delta R=R(2)-R_L(2)$; $P_{1}$ is the inner binary period.
The $R(2)$ distance needs to be calculated carefully since the equation for $\dot{M}(2)$ is very sensitive to the amount of overfill. We decided to adjust $R(2)$ to obtain the $\dot{M}(2)$ value that we report here in Table \ref{tab:initial}; in Fig. \ref{fig:LUCamelexplain} the value of $R(2)$ is represented by a purple horizontal line. Since $R_{L}(2)$ is a function of time, instead of using that, we use the mean value of it, $R_{L}(2)_{mean}$, shown as a green line. Hence we adjust the value $R(2)$ for each integration (in Figure \ref{fig:LUCamelexplain} the system is {\sl LU Cam}), until the difference given by $\Delta R=R(2)-R_L(2)_{mean}$ is the right one, such that $\log \dot{M}(2)$ is as in Table~1.

We can calculate the maximum and minimum of the mass transfer rate by using the values of $R_L(2)_{max}$ and $R_L(2)_{min}$ to obtain $\dot{M}(2)_{max}$ and $\dot{M}(2)_{min}$.

There are two main sources to CV\textquotesingle s luminosity; the hot spot and the disc. The luminosity due to the so--called hot spot is produced when a stream of stellar mass crosses the $L_1$ point and collides with the disc; its expression (Warner~1995) is given by: 
\begin{equation}
L(SP) \approx {G M(1) \dot{M}(2) \over r_{d}},
\end{equation}
where $L(SP)$ is the luminosity due to the hot spot, the radius of the disc is typically $r_{d} \approx 0.40\times a_{1}$ with $a_{1}$ being the semi--major axis of the inner binary (see Table~\ref{tab:initial}).
Applying this equation to our extreme values on $R_L(2)$ we obtain the $L(SP)_{max}$ and $L(SP)_{min}$ values.

Alternatively, the luminosity due to the accretion disc using Eq. 2.22a of Warner~(1995), is:
\begin{equation}
L(d)\approx {1 \over 2} {G M(1) \dot{M}(2) \over R_{1}}.
\end{equation}
Using this equation we can obtain the extreme values of $L(d)_{max}$ and $L(d)_{min} $ for each system. The total luminosity for each extreme is found by adding the estimated luminosity of the hot spot plus the luminosity of the disc, obtaining $L(d)_{T_{max}}$ and $L(d)_{T_{min}}$ for each system.

Then, it is possible to calculate the bolometric magnitude using $M_{bol}=-2.5 \log (L/L_0)$, with $L_{0}=3.0128 \times 10^{28}$ Watts used as a standard luminosity for  comparison. From the extreme values we obtained  $M_{B_{max}}$ and $M_{B{min}}$, leading to a magnitude difference  $\Delta M_{B}$.

\subsection{Magnetic case}

{\sl V1007 Her is the only magnetic system among our selection,} which according to Wu~\&~Kiss~(2008) is a polar system. The accretion luminosity of an accreting white dwarf is given by:
\begin{equation}
\label{eq:Lpolar}
L_{acc} = - {G M(1) \dot{M}(2) \over R_{1}}.
\end{equation}

\begin{table}
\caption{\label{tab:GR} GR periods for all systems obtained using the first order post--Newtonian correction. }
\begin{tabular}{@{}lrrr}
\hline
 Name of the CV& VLPP  & GR period & GR period \\
  & (days) & (days)  & (years) \\
 \hline
 \hline
{\sl LU Camelopardalis} & 265.76  & 27851.13 & 76.25  \\
{\sl QZ Serpentis} & 277.72  & 11445.08 & 31.33 \\
{\sl V1007 Her} & 170.59  & 11245.42 & 30.79 \\
{\sl BK Lyncis} & 42.05  & 9626.77 & 26.36 \\
\hline
\end{tabular} \\
\end{table}

\begin{table}
\fontsize{9}{10}\selectfont
\caption{\label{tab:brightness} Summary of values used to estimate the integration that best fits the VLPP and the change of magnitude for each system.}
\begin{tabular}{@{}lcccc}
\hline
Variable & {\sl LU Cam} & {\sl QZ Serp} & {\sl V1007 Her} & {\sl BK Lyn} \\
 \hline
 \hline
$P_{2}/P_{1}$ & 7.1 & 12.5 & 13.0 & 5.9 \\
$M_3$ (M$_J$) & 97 & 0.63 & 148 & 88 \\
$a_2$ (AU) & 0.021 & 0.019 & 0.021 & 0.011 \\
$\Delta M_{B}$ & 0.55 & 0.07 & 0.73 & 0.68 \\ 
\hline
\end{tabular} \\
\end{table}

For polars in a high state, $L_{acc}$ is much higher than the intrinsic luminosity of the two stars. Thus, we have $L_{bol}\approx L_{acc}$. Polars are Roche--lobe filling systems, with the mass transfer rate given by Eq. \ref{dotM2}, again using Sepinsky~et~al.~(2007) to calculate $R_{L}(2)$ directly from the integration. Therefore, from Eqs.~\ref{dotM2} and \ref{eq:Lpolar}, it is possible to estimate the change in brightness for {\sl V1007 Her} from $L_{acc\_max}$ and $L_{acc\_min}$.

\section{Results and Discussion} 

We studied an ensemble of initial conditions for a hypothetical third body in each system and the way it affects both the VLPP and the change of brightness. All the results of the numerical integrations are shown as black points in Figs.~1,~2,~3 and 4 which correspond to {\sl LU Cam}, {\sl QZ Serp}, {\sl V1007 Her} and {\sl BK Lyn},  respectively.

The upper panel of each figure shows the resulting secular periods of the binary eccentricity as a function of the mass of the perturber. Each curve corresponds to different  $P_2/P_{1}$ ratios. The  thick horizontal line corresponds to the VLPP value of each system.

For a given $P_2/P_{1}$ ratio (i.e. a given curve), some of our integrations produce secular perturbations as we change the mass of the system that never reach the VLPP line. We argue that only systems that cross the VLPP line can explain the long--term change in the light--curve.

The middle panel is a plot of the perturber's semi-major axis against its mass. The black points denote the results of the numerical integrations, while the solid curves are analytical solutions from Georgakarakos~(2009) ($e_2=0$, blue curve) and Georgakarakos~(2003) (eccentric cases, green and red curves).  The straight line denotes the orbital stability limit as given in Holman~\&~Wiegert~(1999), while the dotted line is the stability limit based on the Georgakarakos (2013) results.  In contrast to Holman \& Wiegert~(1999), Georgakarakos~(2013) does not assume a massless particle for any of the three bodies. Hence, two branches of the dotted line are due to the dependence of the stability limit on the mass of the perturber.

\begin{figure}
\begin{center}
\includegraphics[width=9cm]{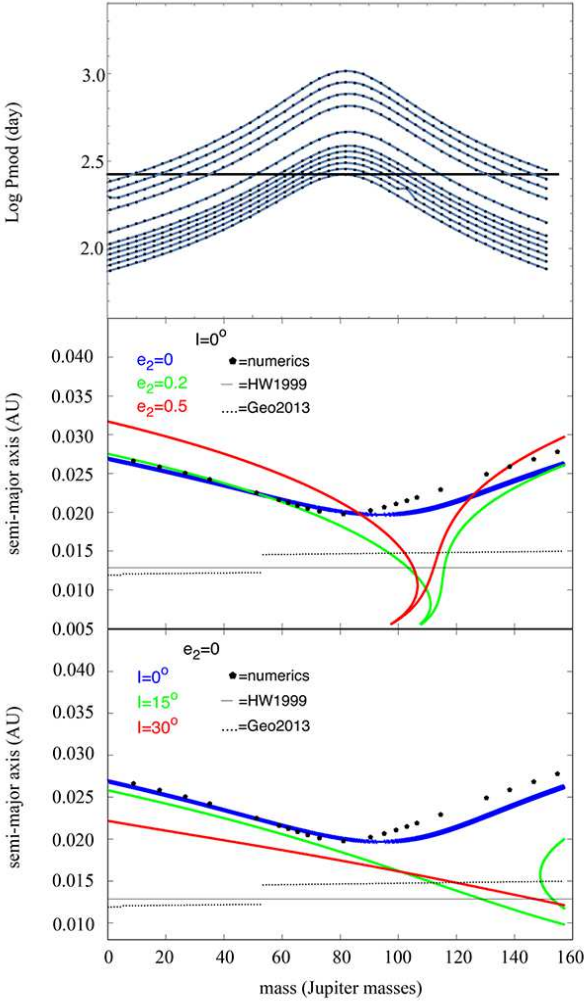}
\vspace{-0.75cm}
\caption{Results for {\sl LU Camelopardalis} system. Numerical integrations performed are represented by black points. ({\sl Top}) Period of the long--term modulation (secular period) as a function of the third-body mass. Each blue curve joining black points correspond to different $P_2 / P_1$ ratios. The black line around 2.4 corresponds to the observed VLPP. Only numerical integrations that can explain the observed VLPP are shown ({\sl middle}) . The blue curve corresponds to the planar and circular planar analytical solution. The green line represents the analytical planar systems with eccentricity of 0.2 and the red line represents the planar systems with eccentricity of 0.5. The doted line represents the inner stability limit calculated by Georgakarakos~(2013) and the grey solid line that of Holman \& Wiegert~(1999). ({\sl Bottom}) Similar quantities as in middle panel, but for a circular orbit with different inclinations. The third body values consistent with observations obtained here are: $M_3=97\, M_J$ and $P_2=1.06$~days.
}
\label{fig:LUCamel}
\end{center}
\end{figure}

\begin{figure}
\begin{center}
\includegraphics[width=9cm]{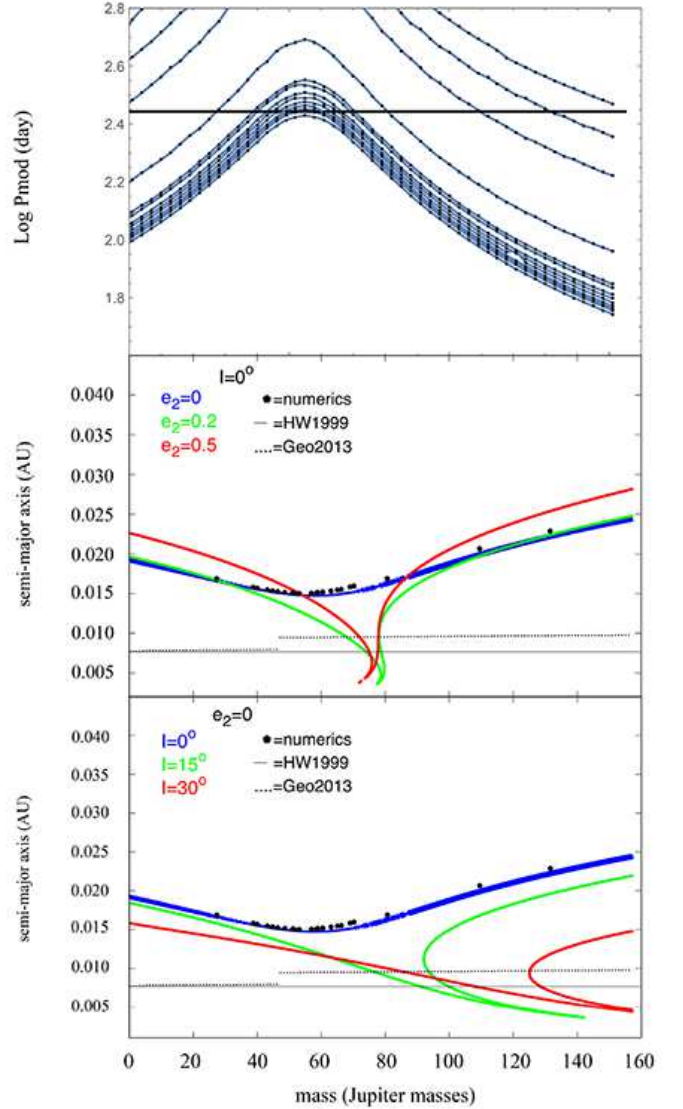}
\vspace{-0.75cm}
\caption{Results for {\sl QZ Serpentis} system. In this system the third body is found to have a mass  $M_{3} = 0.63  \textrm{M}_{\textrm{J}}$ and $P_{2}=1.04$ days.
}
\label{fig:QZSer}
\end{center}
\end{figure}

\begin{figure}
\begin{center}
\includegraphics[width=9cm]{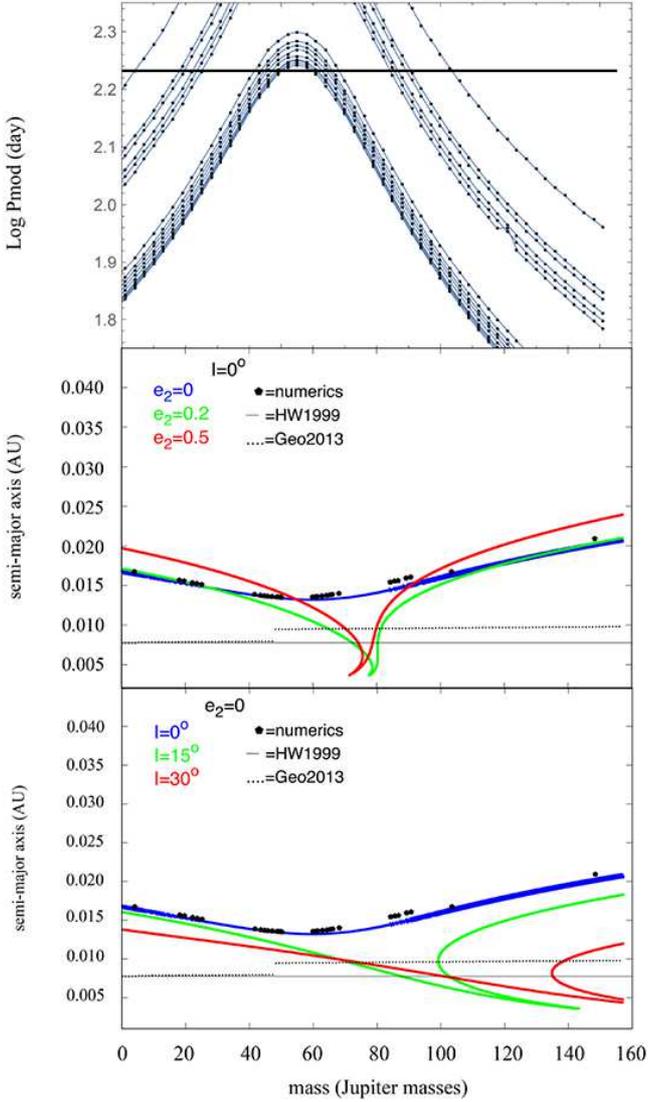}
\vspace{-0.75cm}
\caption{Results for the {\sl V1007 Herculis} system. Integrations yield for the third body a mass of $M_{3} = 148  \textrm{M}_{\textrm{J}}$ and $P_{2}=1.08$ days. }
\label{fig:V1007}
\end{center}
\end{figure}

The lower panels in Figures  \ref{fig:LUCamel}--\ref{fig:BKLyn} are similar to what we presented in the middle panel, but the inclination is varied here. For the coplanar case (blue curve) we use Georgakarakos~(2009), while for the three dimensional cases (green and red curve), we make use of Georgakarakos~(2004).

Table~\ref{tab:brightness} lists the ratio between the period of the third body compared to the period of the inner binary (that is $P_{2} / P_{1} $), the mass of the third body (in Jupiter masses ${\rm M}_J$), and the semi-major axis of the third body ($a_2$ in AU) and the change of magnitude of each system ($\Delta M_{B}$).  We can compare the magnitude change for each system to the observed ones that appear in Table~\ref{tab:initial}.

Now we will discuss some details of the results for each CV. 
We searched for all the numerical integrations whose secular period matched the observed period of the system, and then made all the required calculations in order to estimate the change in magnitude that arises from the perturbations of the third body. A search was done until a system was found that matched the observed change of magnitude of the system. It led to a system that can simultaneously explains the VLPP and the change in magnitude.

\subsection{\sl LU Camelopardis}

This CV has an observed VLPP of 265.76~days, with ${M_{2} / M_{1}} = 0.34$, which is the largest ratio among the CVs studied here. Figure~\ref{fig:LUCamel} shows our numerical results for this system.

The stability limit given by Holman~\&~Wiegert~(1999)  and Georgakarakos~(2013) are also shown. Holman \&
 Wiegert rule out any $a< 0.013$~AU (grey horizontal line), while Georgakarakos rules out any $a< 0.015$~AU 
(black dashed line). Care has to be exercised in the eccentric cases when dealing with small values of the semi-major
axis since the analytical formulae have singularities, This holds for the rest
 of the systems.

In this particular system the third body initially is on a circular orbit that explains the observational value $P_{2} / P_{1}  = 7.1$, that is $P_{2}=25.5$ h = 1.06 days, its mass being $M_{3} = 97 \, \textrm{M}_{\textrm{J}}$, and a semi--major axis of $a_{3}=0.021$~AU. These system parameters also match the observed $\Delta M_{B}=0.55$.

Alternatively, the long period calculated using first--order GR correction for this systems is 27851.13 days (76.25 years), which is far too large to explain the observed VLPP of 265.76 days.

\subsection{\sl QZ Serpentis}

The VLPP observed for this system is 278~days, the mass ratio is $ {M_{2} / M_{1}} = 0.20$. Figure~\ref{fig:QZSer} shows our numerical and analytical results for the circular and eccentric conditions.
Here, Holman \& Wiegert (1999) rule out any $a< 0.0075$~AU (grey horizontal line).  Alternatively, Georgakarakos~(2013) rules out $a< 0.010$~AU (black dashed line).
Recall that some singularities can appear for small values of $a$.

In this particular system the third body that is initially on a circular orbit and explains the observed
  $P_{2} / P_{1}  = 12.5$, that is $P_{2}=24.9$ h = 1.04 days, with its mass being $M_{3} = 0.63 \, \textrm{M}_{\textrm{J}}$ (the smallest among the systems), and a semi--major axis of $a_{2}=0.019$~AU. This system matches the observed $\Delta M_{B}=0.07$.

In a similar fashion, for the inclined orbits with $i=15^{\circ}$ we observed that the values in {$a$} get higher than in the circular case, but at a faster rate than when exploring the eccentric cases. For $i=30^{\circ}$ the masses increase faster than the circular case as we decrease the semi--major axis $a$. 

The GR first order correction for this system yields a period of 11445.08 days (31.33 years), far too large to explain the 277.72 days of the observed period.

\begin{figure}
\begin{center}
\includegraphics[width=9cm]{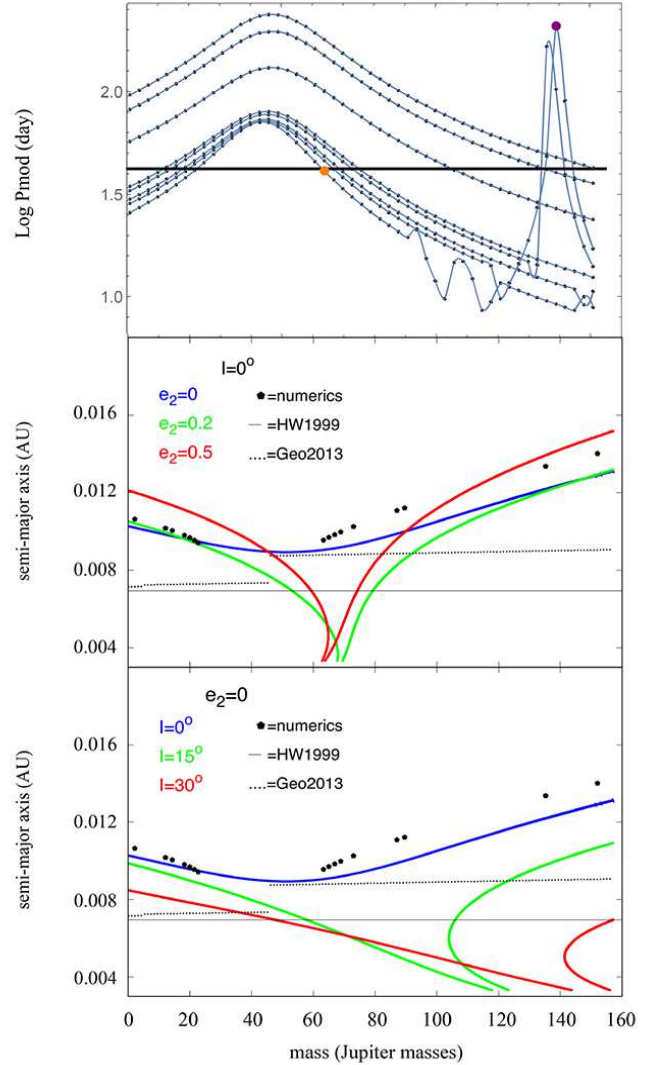}
\vspace{-0.75cm}
\caption{Results for the {\sl BK Lyncis} system. Here, the third body has a mass of $M_{3} = 88  \textrm{M}_{\textrm{J}}$ and period of $P_{2}=0.44$ days.
}
\label{fig:BKLyn}
\end{center}
\end{figure}

\begin{figure*}
\begin{center}
\includegraphics[width=10cm,clip=20cm]{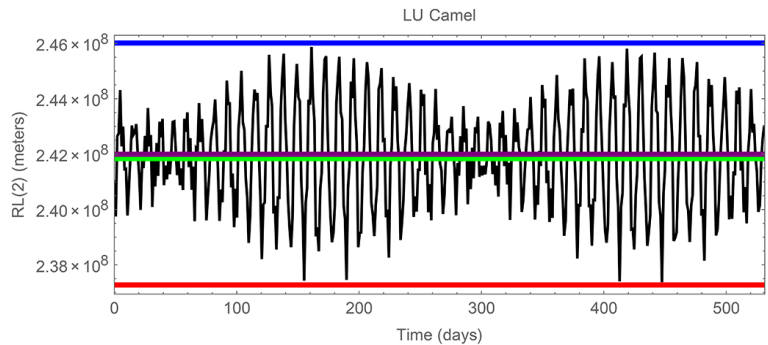}
\caption{Method used to calculate the change of magnitude due to the third body. The time evolution of $R_{L}(2)$ for the CV {\sl LU Cam} is shown as an example. The blue horizontal line is the maximum value for the  $R_{L}(2)$  that the system reaches ($R_L(2)_{max}$), the red one corresponds to the minimum value ($R_L(2)_{min}$), green is the mean value ($R_L(2)_{mean}$) and purple is the $R(2)$ value. See text for more details.}
\label{fig:LUCamelexplain}
\end{center}
\end{figure*}

\subsection{\sl V1007 Herculis}

This CV has an observed VLPP of 170.59~days, and the mass ratio is given by $M_{2} / M_{1} = 0.202$. Figure~\ref{fig:V1007} shows the result of the numerical integrations performed, including circular (numerical integrations) and eccentric (analytical) orbits as well as two cases with different inclinations (analytical) as we did before.

Holman \& Wiegert~(1999) rule out any $a< 0.0075$~AU (grey horizontal line), and  Georgakarakos~(2013) rules out any $a< 0.010$~AU (black dashed line).

In this particular system the third body initially on a circular orbit that give us a closer value to what we observe is $P_{2} / P_{1}  = 13.0$ (i.e. $P_{2}=25.98$ h = 1.08 days), with mass of $M_{3} = 148 \,  \textrm{M}_{\textrm{J}}$, and a semi--major axis of $a_{2}=0.021$~AU. This system  does not match the observed $\Delta M_{B}=0.97$ but is the closest to that value with $\Delta M_{B}=0.73$.

Lastly, the GR post Newtonian first--order correction give us a period of 11245.42 days (or 30.788 years), which can not explain the VLPP of 170.59 days.

 \subsection{\sl BK Lyncis}

The observed VLPP is 42.05~days and $M_2/M_1=0.1674$, with both values are being the lowest observed in our set of CVs. Similarly to the previous figures, Fig.~\ref{fig:BKLyn} shows our results for {\sl BK Lyncis}.

The empirical criterion of Holman \& Wiegert~(1999) rules out any $a< 0.007$~AU (grey horizontal line), while the work of Georgakarakos~(2013) implies that $a>0.009$~AU  (black dashed line) can be ruled out.

In this system, since both stability limits are higher than the minimum of the curve for the initially circular--planar case (black dots and blue curve on Figure 5 middle and lower panel), the possible solutions for this system are likely to be higher than those limits.   In fact, we could not find numerically (black dots) stable orbits below 0.0094~AU; this will be discussed in the following section.

For {\sl BK Lyncis} the third body on an initially circular--planar orbit that best reproduces what we observed in the light curve of the CV has $P_{2} / P_{1}  = 5.9$ ($P_{2}=10.62$~h = 0.44~days), mass $M_{3} = 88 \, \textrm{M}_{\textrm{J}}$, and  $a_{2}=0.011$~AU. This system matches the observed $\Delta M_{B}=0.68$.

The GR first--order correction can not explain the observed VLPP period of 42.05 days, since the predicted period is 9626.77 days (26.36 years), which is far too large.

\section{Final Comments}

In this article we explored the possible origin of the very long photometrical periods (VLPPs) observed in four cataclysmic variables; {\sl LU Camelopardalis}, {\sl QZ Sepentis}, {\sl V1007 Herculis}, {\sl BK Lyncis}, all of them first reported by Yang~et~al.~(2017).

We find that three out of four of the systems can be explained by the secular perturbations of a third body orbiting around each CV. In the case of {\sl V1007 Herculis} we could not find an initially circular planar orbit that could explain the relatively large change in magnitude observed $\Delta M_{B} \approx 1$.

All of our numerical integrations and modelling are based on the parameters estimations calculated using Knigge~et~al.~(2011), and using the orbital period of the CV as the starting point. Then, we estimated the best dynamical parameters of each system.

{\sl Lu Camelopardalis} was explored assuming initially circular planar orbit by numerical means and found that the configuration that explains both the observed VLPP and change of magnitude $\Delta M_{B}$ has a period of $P_{2} = 25.5$ h, and a mass of $M_{3} = 97 \, \textrm{M}_{\textrm{J}}$, that is larger than the minimum mass required for having nuclear reactions at its centre (83~M$_{J}$).

Alternatively, for {\sl QZ Sepentis} the configuration that can explain both the observed VLPP and change of magnitude $\Delta M_{B}$ has a $P_{2} = 24.9$ h and $M_{3} = 0.63~\textrm{M}_{\textrm{J}}$ . This third body mass is small compared to the rest of the CVs and is well within the planetary mass. These results are most likely because the observed change of magnitude is quite small ($\Delta M_{B}=0.07$).

In {\sl V1007 Herculis} the third body that best fits the observed VLPP and the change of magnitude has $P_{2} = 25.98$ h and $M_{3} = 148\,  \textrm{M}_{\textrm{J}}=0.141 \, {\rm M}_\odot$; this system produces a $\Delta M_{B}=0.73$. This mass is far too big (as the mass of a red dwarf) and even using this high value it was not possible to reproduce the observed change in magnitude of $\Delta M_{B}=0.97$. We conclude that, if all our estimations are correct, it is only marginally possible that a third body in a close to circular and near planar orbit can explain the VLPP and change of magnitude on this system.

\begin{figure}
\begin{center}
\includegraphics[width=8.5cm]{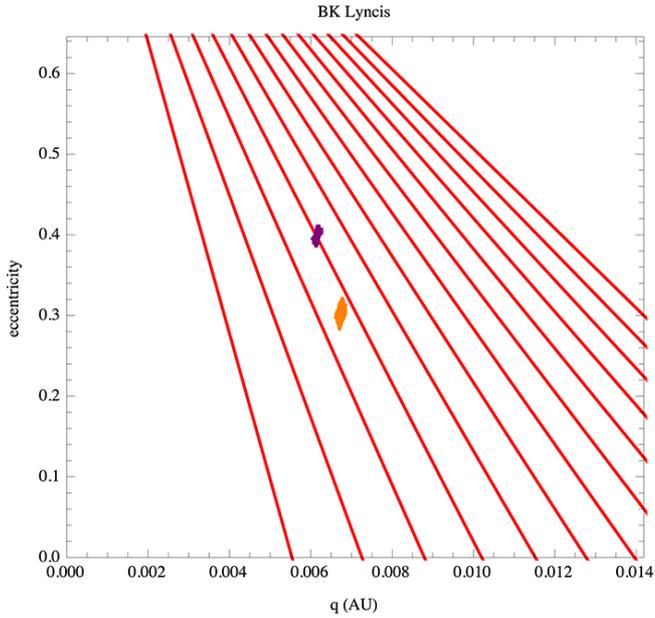}
\vspace{-0.55cm}
\caption{Search for mean motion resonances in the {\sl BK Lyncis} system. Eccentricity as a function of the pericentre distance $q(AU)$ is shown. The time evolution of two systems is  displayed: one in resonance (purple) and the other not (orange). 
The lines in red, from left to right, represent the 2:1, 3:1, 4:1, 5:1, 6:1, 7:1, 8:1, 9:1, 10:1, 11:1, 12:1, 13:1 and 14:1 mean motion resonances between the inner binary and the third body. See text for discussion.
}
\label{fig:BKLynRes}
\end{center}
\end{figure}

Lastly, we have the system {\sl BK Lyn} which was the most challenging one to model. First we realized that there are not stable systems with orbits below 0.0094~AU. Then we could not find the full curve for the numerical integrations in Figs.~\ref{fig:BKLyn} (middle and bottom plots). The initially circular analytical curve (shown in blue) helps as a reference of how the full curve would look like. 
It is important to point out that even when the analytical curves are very helpful to see what the effect of the eccentricity and inclination have on the VLPP, we do not know which orbits are stable or not.
   The system that can explain both the observed VLPP and change of magnitude $\Delta M_{B}$ has $P_{2} = 5.9$ h and $M_{3} = 88 \,  \textrm{M}_{\textrm{J}}$ which is close to the mass threshold of $83\,  \textrm{M}_{\textrm{J}}$  of becoming an M star. 
   Also in this system we noticed that in Fig.~\ref{fig:BKLyn} top panel, the curves for low $P_{2} / P_{1}$ (that is the four curves at the bottom of the plot), when they are close to 90--100 $ \textrm{M}_{\textrm{J}}$ their behaviour changes and seem to oscillate and have an abrupt increase in the value of Pmod (y--axis). After careful consideration, we realize that those sudden changes must originate from mean motion resonances.
    Fig.~\ref{fig:BKLynRes} shows the eccentricity as a function of pericentre distance for two systems. The orange dots correspond to $M_{3} = 63 \, \textrm{M}_{\textrm{J}}$, $a=0.0062$~AU and mean eccentricity of $e=0.399$. The purple ones  correspond to a system with $M_{3} = 139 \, \textrm{M}_{\textrm{J}}$,  $a=0.0067$~AU and $e=0.302$. 
   These initial conditions were chosen to provide an example of a system that evolves due to secular perturbations (orange) and another that evolves due to resonant perturbations (purple). It is observed from Fig.~\ref{fig:BKLynRes} that the resonant system (purple) is immersed in the 5:1 Mean Motion Resonance (MMR) while the orange system is between the 4:1 and 5:1 MMR.

   We also show in Figure~\ref{fig:BKLyn} top plot where the purple and orange systems (as circles with their corresponding color) are located. This indicates that the peaks observed in the low-right part of the plot are indeed due to resonances.

   As it can be appreciated in Fig.~\ref{fig:BKLyn} top plot, the curves associated with resonances increase their Pmod value very quickly as the system approaches the resonance (crossing the observational VLPP black line). Then it is possible to find configuration families similar to Fig.~\ref{fig:BKLyn} middle and bottom plots that can explain the observed VLPP but using resonant systems, instead of the secular families that we studied here. We decided to leave the study of the resonant families for a future contribution.

   We find that for all systems the first--order post Newtonian GR corrections cannot account for the observed VLPP in any of the systems, since the predicted periods are far too large compared to the observed ones.

   All the parameters of the pair of stars (masses, semi--major axes, radii, etc)  that form the CVs were estimated using the orbital period observed. Since these are based on average statistical values, our estimations on the parameters of the third body, rather than being precise values, are also estimates for the possible third body that might explain the observed characteristics.

   In three out of the four systems we found a third body on an initially circular orbit that explains both the observed VLPP and change in magnitude by secularly perturbing the inner binary. 
In the case of {\sl V1007 Herculis} it was not possible to find a numerical model that could account for the $\sim 1$ change in magnitude. 
We also find that for {\sl BK Lyncis} mean motion resonances are important, hence  it  is possible that a third body in resonance could also account for the observed VLPP. Further exploration of the role of resonances in these systems is postponed for future works, since here we focused on secular perturbations.

\section*{Acknowledgements}
We thank the referee for the useful comments and corrections.
CEC would like to thank IAChR and JRChR for their helpful discussions and to WBRA  for her advises and help on the development of this article.

GT acknowledges support from PAPIIT project IN110619.

\section*{Data Availability}
The data presented and discussed in this article will be shared on reasonable request to the corresponding author.






\appendix

\section*{Appendix}
Secular period for coplanar orbits, low eccentricity perturber (see Georgakarakos~2009 for more details):
\begin{equation}
T_s=\frac{2\pi}{(k_3-k_4)\tau},
\end{equation}
where
\begin{equation}
k_{3}=\sqrt{\frac{k_{1}+k_{2}}{2}},\hspace{0.1cm}
\quad 
k_{4}=\sqrt{\frac{k_{1}-k_{2}}{2}},\hspace{0.1cm}
\end{equation}
\begin{equation}
k_{1}=D^{2}+B^{2}+2C^{2}D, 
\quad
k_{2}=\sqrt{(D+B)^{2}[(D-B)^{2}+4C^{2}D]},
\end{equation}
\begin{equation}
B=1+\frac{75}{8}\gamma, \quad C=\frac{5}{4}\alpha
\quad \mbox{and} \quad D=\beta,
\end{equation}

\begin{equation}
\label{beta}
\alpha =\frac{M_1-M_2}{M_1+M_2}\bigg(\frac{a_1}{a_2}\bigg),  
\qquad
\beta
=\frac{M_1M_2M^{\frac{1}{2}}}{M_3(M_1+M_2)^{\frac{3}{2}}}\bigg(\frac{a_1}{a_2}\bigg)^{\frac{1}{2}},
\end{equation}

\begin{equation}
\label{gamma}
\gamma=\frac{M_3}{M_T^{\frac{1}{2}}(M_1+M_2)^{\frac{1}{2}}}\bigg(\frac{a_1}{a_2}\bigg)^{\frac{3}{2}}
\end{equation}
and
\begin{equation}
\label{tau}
\tau=\frac{3}{4}\frac{G^{\frac{1}{2}}M_3a^{\frac{3}{2}}_1}{a^{3}_2(M_{1}+M_{2})^{\frac{1}{2}}}.
\end{equation}
Here $M_T$ is total mass of the system.

The secular period for coplanar orbits, eccentric perturber (see Georgakarakos~2003 for more details):
\begin{equation}
T_s=\frac{2\pi}{|(B-A)\tau|},
\end{equation}
where
\begin{displaymath}
A=\frac{\beta}{(1-e^{2}_2)^{2}},\hspace{0.3cm} B=\frac{1}{(1-e^2_2)^{\frac{3}{2}}}+\frac{25}{8}\gamma\frac{3+2e^2_2}{(1-e^2_2)^3}
\end{displaymath}
and $\beta$, $\gamma$ and $\tau$ are given by eqs (\ref{beta}), (\ref{gamma}) and (\ref{tau}).

Secular period for non-coplanar orbits, low eccentricity perturber (see Georgakarakos~2004 for more details):
\begin{equation}
T_s=\frac{2\pi}{|(\sqrt{BD}-A)\tau|},
\end{equation}
where
$$
A=\cos{i}+\frac{1}{2}\beta(4-5\sin^{2}{i}) \;,  
$$
$$
B=2-5\sin^{2}{i}+\beta\cos{i} \;, \quad
D=2+\beta\cos{i}\;, 
$$
with $i$ the mutual inclination and $\beta$ and $\tau$ are given by Eqs.~(\ref{beta}) and (\ref{tau}), respectively.


\bsp	
\label{lastpage}
\end{document}